\documentclass[aip,jcp,reprint,longbibliography]{revtex4-2}

\usepackage{graphicx}
\usepackage{amsmath}
\usepackage{amssymb}
\usepackage[colorlinks=true,allcolors=blue]{hyperref}
\usepackage{xcolor}

\newcommand{\R}{{\mathcal{R}}}

\begin{document}

\title{Continuous-time multifarious systems - Part II: non-reciprocal multifarious self-organization}

\author{Jakob Metson}
\affiliation{Max Planck Institute for Dynamics and Self-Organization (MPI-DS), 37077 G\"ottingen, Germany}

\author{Saeed Osat}
\email{saeedosat13@gmail.com}
\affiliation{Max Planck Institute for Dynamics and Self-Organization (MPI-DS), 37077 G\"ottingen, Germany}

\author{Ramin Golestanian}
\email{ramin.golestanian@ds.mpg.de}
\affiliation{Max Planck Institute for Dynamics and Self-Organization (MPI-DS), 37077 G\"ottingen, Germany}
\affiliation{Rudolf Peierls Centre for Theoretical Physics, University of Oxford, Oxford OX1 3PU, United Kingdom}

\date{\today}

\begin{abstract}
In the context of self-assembly, where complex structures can be assembled from smaller units, it is desirable to devise strategies towards disassembly and reassembly processes that reuse the constituent parts. A non-reciprocal multifarious self-organization strategy has been recently introduced, and shown to have the capacity to exhibit this complex property. In this work, we study the model using continuous-time Gillespie simulations, and compare the results against discrete-time Monte Carlo simulations investigated previously. Furthermore, using the continuous-time simulations we explore important features in our system, namely, the nucleation time and interface growth velocity, which comprise the timescale of shape-shifting. We develop analytical calculations for the associated timescales and compare the results to those measured in simulations, allowing us to pin down the key mechanisms behind the observed timescales at different parameter values.
\end{abstract}


\maketitle

\section{Introduction}

The natural world is filled with intricate structures that emerge from the self-assembly of simpler building blocks, ranging from the folding of proteins to the formation of cellular membranes. Harnessing these processes for synthetic materials has been a longstanding goal\cite{Mendes2013WNN,Bassani2024}, as it opens avenues for creating systems capable of dynamic adaptation. 
For instance, multifarious self-assembly systems can be designed to assemble multiple distinct target structures from a shared pool of building blocks \cite{Murugan2015PNAS, Sartori2020PNAS, Osat2023NN,Bohlin2023AN,Osat2024PRL,Evans2024N}. This is an important concept in natural self-assembly. For example, the myriad of assembled proteins are all formed from the same set of around twenty amino acids.
In recent years, research into self-organizing systems has revealed that incorporating mechanisms for disassembly and reassembly can further enhance the complexity and functionality of these materials.
Many mechanisms for dynamic self-organization have been explored\cite{Nguyen2011AN, Shen2024NN, Sarraf2023SR, Herron2023PL, Wang2024NJP, Landi2025SM}, including allosteric interactions \cite{Zhang2017NC, Zhu2024PRR, Metson2025PRR, Evans2024JCP} and modifying the chemical potential of tiles \cite{Bisker2018PNAS, Faran2023JPCB, Faran2025JCIM}.
One promising approach in this domain involves non-reciprocal interactions, which break traditional symmetry in interactions and allow for more versatile self-organization \cite{Soto2014,Soto2015,AgudoCanalejo2019,Saha2019,Saha2020PRX,OuazanReboul2021,OuazanReboul2023b,OuazanReboul2023,OuazanReboul2023a,Osat2023NN,Parkavousi2025PRL,Golestanian2024EN}.

\begin{figure}[b]
\begin{center}
\includegraphics[width=0.9\linewidth]{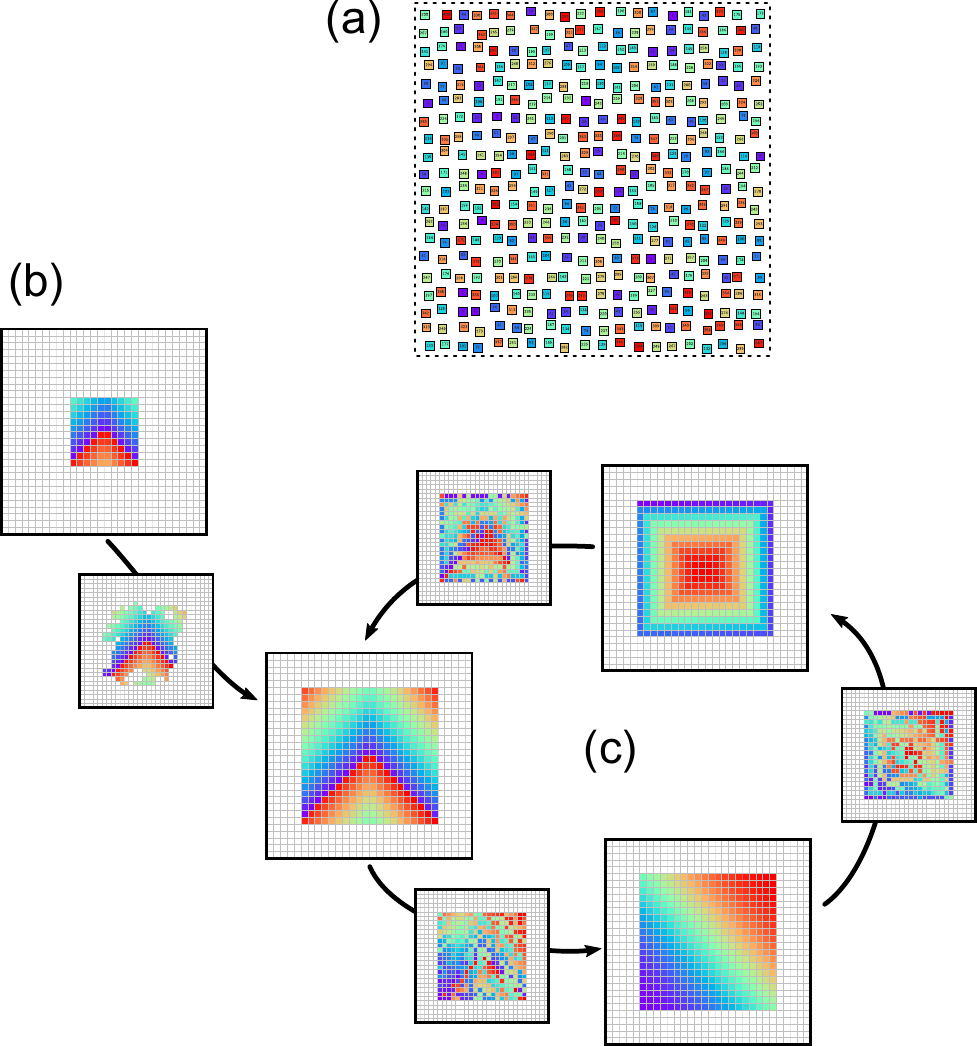}
\caption{An illustration of non-reciprocal multifarious self-organization. Three different structures correspond to particular arrangements of the different tile species shown in (a). (b) By placing an appropriate seed into the system, a desired structure can be retrieved and assembled. (c) The system can then autonomously rearrange the components to `shape-shift' between structures.
\label{fig:Cartoon}}
\end{center}
\end{figure}

In many traditional self-assembly models, components interact reciprocally, meaning the interaction between any two components is equal and opposite. While this reciprocity is fundamental to many physical systems, the introduction of non-reciprocal interactions offers a way to bypass the limitations of equilibrium states. Non-reciprocal interactions\cite{Golestanian2024EN} can drive systems out of equilibrium, leading to the emergence of new dynamic phases and structures that are not possible in reciprocal systems\cite{Soto2014,Ivlev2015PRX,Fruchart2021N}. For example, non-reciprocal multi-component scalar field theories have revealed novel phases, such as chasing bands and active lattice-like phase separations \cite{Saha2020PRX,You2020PNAS, Pisegna2024PNAS,Brauns2024PRX, Rana2024PRL, Parkavousi2025PRL}. Non-reciprocal interactions have been identified in various physical and biological systems, often resulting in unexpected and fascinating behaviors. Examples include non-reciprocal couplings between chemically interacting particles \cite{Soto2014, Soto2015, AgudoCanalejo2019, Saha2019, Mandal2024C} with strong implication on self-organization of metabolic cycles at the origins of life\cite{OuazanReboul2021,OuazanReboul2023b,OuazanReboul2023,OuazanReboul2023a}, quorum-sensing active matter \cite{Duan2023, Dinelli2023NC,Duan2025}, ecological populations \cite{Blumenthal2024PRL, Marcus2024PRE}, active matter with hydrodynamic interactions \cite{Uchida2010PRL, Hickey2023PNAS,Golestanian2025}, spin models \cite{Uchida2010,Loos2023PRL, Hanai2024PRX, Blom2025PRE, Saha2024arXiv}, and synthetic systems with engineered interactions \cite{Sharan2023,Tucci2024,Brandenbourger2019NC}. These findings underscore the pervasive nature of non-reciprocal interactions, alongside their potential for producing complex behaviors. The versatility of non-reciprocal interactions suggests they could serve as a foundational principle in designing adaptive and programmable materials.

Combining non-reciprocal interactions with multifarious self-assembly, the same set of components can rearrange into different configurations, facilitating dynamic transitions between multiple target structures\cite{Osat2023NN}.
In this study, we focus on the dynamics of non-reciprocal multifarious self-organization. Building upon previous works \cite{Osat2023NN, Osat2024PRL} that employed discrete-time Monte Carlo simulations, we investigate the behavior of the model using continuous-time Gillespie simulations. This allows us to validate that both simulation methods give results that are largely in agreement. In addition, using the continuous-time simulations we study key timescales, such as nucleation time and interface growth velocity. By measuring these timescales and comparing them with theoretical calculations, we identify the primary mechanisms that govern shape-shifting behavior under different parameter values.
Our investigation examines how non-reciprocal interactions drive the transition between structures, exploring the stability of different growth mechanisms. We investigate how factors such as interaction strength, non-equilibrium drive, and chemical potential influence the kinetics of the shape-shifting process. Through this study, we seek to deepen our understanding of non-reciprocal self-organization, potentially guiding the development of advanced synthetic systems capable of dynamic reconfiguration.

\begin{figure}[t]
\begin{center}
\includegraphics[width=0.7\linewidth]{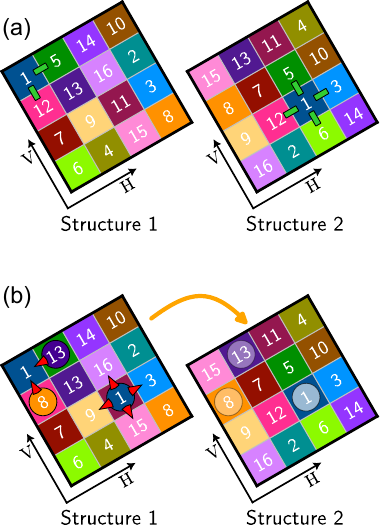}
\caption{All the (a) reciprocal and (b) non-reciprocal interactions involving tile species 1 for an example system storing two $4\times4$ structures with a shifting sequence from Structure 1 to Structure 2. The green rectangular markers show the reciprocal interactions, and the red triangular markers show the non-reciprocal interactions.
\label{fig:Interactions}}
\end{center}
\end{figure}

\section{Model} \label{sec:Model}
Multifarious systems are designed to assemble multiple different structures using a shared set of components. In the equilibrium multifarious self-assembly model \cite{Murugan2015PNAS}, structures are encoded within the interactions between building blocks - a molecular analogy of a Hopfield network \cite{Hopfield1982PNAS}. A structure can be `retrieved' using an appropriate seed, upon which the system assembles the full corresponding structure.

Including non-reciprocal interactions between the building blocks, where the interactions are not symmetric, breaks detailed balance. This leads to a fundamentally non-equilibrium model. The resulting non-reciprocal multifarious self-organization model exhibits a variety of non-trivial behaviors. 
For example, systems can be designed to shape-shift, with assembled structures autonomously reorganizing into new structures, following a custom predefined sequence of structures.
The non-reciprocal interactions can also be used to drive systems out of otherwise extremely long-lived kinetic traps \cite{Osat2024PRL}.

To form the non-reciprocal multifarious self-organization model, we build upon the equilibrium model described in the previous work of this series \cite{Metson2025arXiv-a}, adding non-reciprocal interactions to give a non-equilibrium model.
We design systems to store $m$ structures with dimensions $l\times l$. The structures are assembled in a square lattice of size $L\times L$, which can have either periodic or hard-wall boundary conditions. Each lattice site is in contact with a large reservoir, which contains many copies of each tile species. We use $M=l^2$ different tile species and define the structures as different random arrangements of the tile species, with each structure containing one copy of each tile. Furthermore, through the addition of non-reciprocal interactions, we can design systems to autonomously reorganize, shape-shifting between the structures in predefined sequences. The assembly and shifting behaviors are illustrated in Fig.~\ref{fig:Cartoon}.

The reciprocal interactions represent bonds that form between neighboring tiles, lowering the system energy. 
They are constructed such that the system is able to assemble the different structures encoded in the system given a suitable seed. To do this, each tile species interacts reciprocally with strength $\varepsilon$ to any tile species it neighbors in a structure\cite{Metson2025arXiv-a, Osat2024PRL, Osat2023NN}. 
We denote the set of all specific interactions imposed by the $m$ structures $S^{(1)},S^{(2)},\dots,S^{(m)}$ stored in the system as ${\cal I}^{\mathrm{r}}$. This is the union of smaller sets, each containing the specific interactions imposed by one of the target structures,
\begin{equation}
    {\cal I}^{\mathrm{r}} \equiv 
I^{\mathrm{r}} (S^1) \cup I^{\mathrm{r}} (S^2) \cup \dots \cup I^{\mathrm{r}} (S^m).
\end{equation}
The smaller sets are defined via 
\begin{align}
    I^{\mathrm{r}} (S^{\ell}) = \bigcup_{\substack{ {\left \langle \alpha,\beta \right\rangle } }} \; 
    S^{\ell}_{\alpha}  \square S^{\ell}_{\beta}.
    \label {eq:interactions}
\end{align}
Here $\alpha$ and $\beta$ are lattice coordinates running over nearest neighbors. A specific interaction belongs to the set ${\cal I}^{\mathrm{r}}$ if at least one target structure favors that interaction. To enable clear notation we rotate the lattice as shown in Fig.~\ref{fig:Interactions}(a). $\square \in \{ \diagdown, \diagup \}$ are the vertical and horizontal directions of the specific interactions. Putting everything together, the interaction matrix is then 
\begin{align}
  U_{A \square B}^{\mathrm{r}} = \begin{cases}
    -\varepsilon, & \text{if } A \square B \in  {\cal I}^{\mathrm{r}} \\
    0, & \text{otherwise}.
  \end{cases}
  \label{eqn:U}
\end{align}
The reciprocal interactions and the chemical potential $\mu$, which is defined later, form the equilibrium total energy
\begin{equation}\label{eq:Hamiltonian}
\mathcal{H} =  \sum_{\langle \alpha, \beta \rangle} U^{\rm r}_{\sigma_\alpha \square \sigma_\beta} 
 - \mu \sum_{\alpha} (1-\delta_{0,\sigma_\alpha}).
\end{equation}
Using this energy in discrete-time Monte Carlo or continuous-time Gillespie simulations would model a system in equilibrium. We now describe how the non-reciprocal interactions are constructed, and the resulting non-equilibrium drive.

\begin{figure*}[t]
\begin{center}
\includegraphics[width=0.85\linewidth]{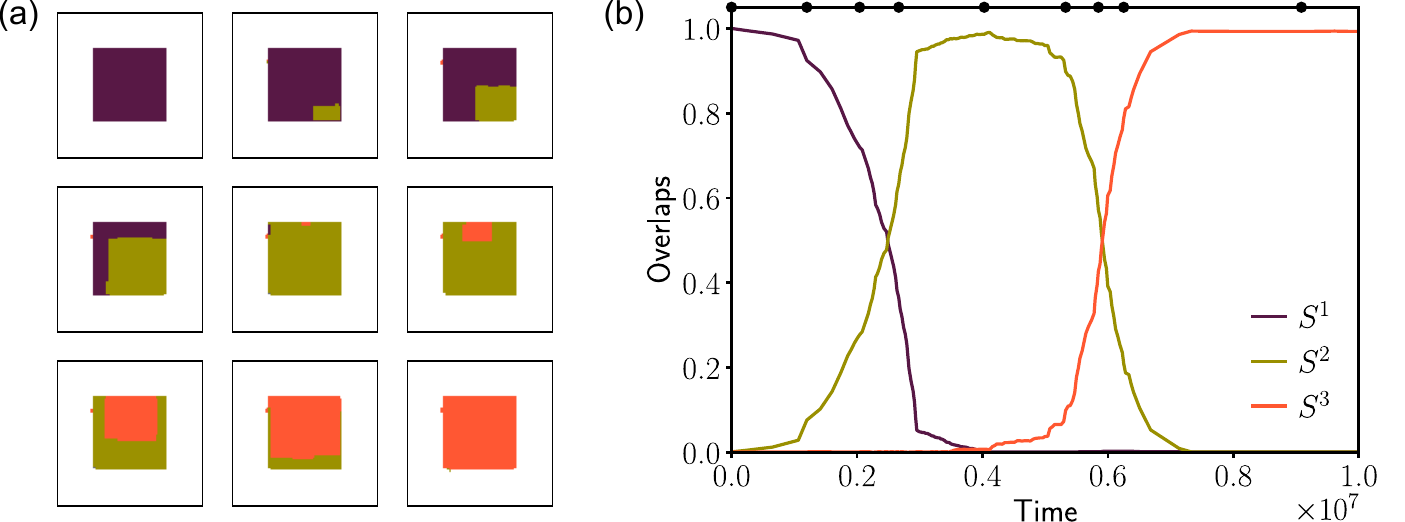}
\caption{(a) Snapshots of shape-shifting for a system with shifting sequence $S^1\to S^2\to S^3$. (b) Evolution of the overlaps (how much of each structure is assembled in the lattice). Black dots along the top axis mark the times of the snapshots shown in (a).
\label{fig:overlaps}}
\end{center}
\end{figure*}

For the non-reciprocal interactions, we use the same procedure as in previous work on non-reciprocal multifarious self-organization \cite{Osat2023NN, Osat2024PRL}. 
The non-reciprocal interactions represent biases in the Monte Carlo acceptance probability and in the Gillespie rates that break detailed balance. 
First, for a given system we impose a shifting sequence $S^1 \to S^2 \to \dots \to S^q$. The non-reciprocal interactions are then designed such that the system is able to shift between these structures in the given sequence. To do this, tiles in a given structure interact non-reciprocally with tiles in the previous structure in the sequence. In particular, they have a non-reciprocal interaction of strength $\lambda$ to tiles they would neighbor if the structure was overlaid onto the preceding structure in the shifting sequence. An illustration of the non-reciprocal interactions is shown in Fig.~\ref{fig:Interactions}(b). For a more precise definition of the non-reciprocal interactions, consider a single shift $S^{\ell} \rightarrow S^{\ell+1}$. A tile will approach and interact with the neighbors of its counterpart in the previous pattern in the sequence. The set of all non-reciprocal interactions imposed for this shift is 
\begin{align}
I^{\mathrm{nr}} (S^{\ell} \rightarrow S^{\ell+1}) = \bigcup_{\substack{ i,j }} \; 
\Big\{ \;
    S^{\ell}_{i-1,j}  \swarrow S^{\ell+1}_{i,j}, \; \nonumber\\
    S^{\ell+1}_{i,j}  \nearrow S^{\ell}_{i+1,j}, \; 
    S^{\ell+1}_{i,j}  \searrow S^{\ell}_{i,j-1}, \;
    S^{\ell}_{i,j+1}  \nwarrow S^{\ell+1}_{i,j}  \;\Big\},
\end{align}
where $(i,j)$ are the indices of the two-dimensional lattice coordinates.
The set of all non-reciprocal interactions for the full sequence is denoted ${\cal I}^{\mathrm{nr}} \equiv 
I^{\mathrm{nr}} (S^1 \rightarrow S^2) \cup \dots \cup I^{\mathrm{nr}} (S^{q-1} \rightarrow S^{q})$. We define
\begin{align}
  R_{A \blacksquare B}^{\mathrm{nr}}= &\begin{cases}
    \lambda, & \text{if } A \blacksquare B \in  {\cal I}^{\mathrm{nr}}, \\
    0, & \text{otherwise},
  \end{cases}
  \label{eqn:R}
\end{align}
where $\blacksquare \in \{ \searrow, \nwarrow, \nearrow, \swarrow \}$ represents all possible specific non-reciprocal interaction directions.
The non-reciprocal interactions give rise to a non-equilibrium drive. For the reaction changing tile $\sigma_{i,j}$ to tile $\sigma'$, at lattice site $(i,j)$, the drive is
\begin{equation}\label{eq:Lambda}
\Lambda = R_{\sigma_{i-1,j} \swarrow \sigma^{\prime}}^{\mathrm{nr}}+
          R_{\sigma^{\prime} \nearrow \sigma_{i+1,j}}^{\mathrm{nr}}+ 
          R_{\sigma^{\prime} \searrow \sigma_{i,j-1}}^{\mathrm{nr}}+
          R_{\sigma_{i,j+1} \nwarrow \sigma^{\prime}}^{\mathrm{nr}}.
\end{equation}
As explained further in Appendix~\ref{app:simulations}, the non-equilibrium drive is used in the Monte Carlo acceptance probability and in the Gillespie rates, breaking detailed balance and resulting in a fundamentally non-equilibrium model.

As discussed in the previous paper \cite{Metson2025arXiv-a}, every lattice site in the system is either empty (occupied by solvent) or occupied by a tile of a particular species.
Each lattice site is connected to a shared large reservoir of solvent and tiles. The concentration $c_i$ of tile species $i$ in the reservoir is obtained by normalizing with respect to the number of solvent tiles in the reservoir $N_0$,
\begin{equation}
    c_i = \frac{N_i}{N_0}.
\end{equation}
From this we obtain the chemical potential of species $i$ as
\begin{equation}
    \mu_i = k_{\rm B} T \log c_i.
\end{equation}
We assume that the reservoir is very large, and set the chemical potential of each tile species to the same constant value $\mu$. By definition the chemical potential of the solvent is 0. 

We use a discrete-time Monte Carlo algorithm and a continuous-time Gillespie algorithm to simulate the time evolution of these models. The details of the simulation methods are presented in Appendix \ref{app:simulations}.

\begin{figure*}[htb]
\begin{center}
\includegraphics[width=\linewidth]{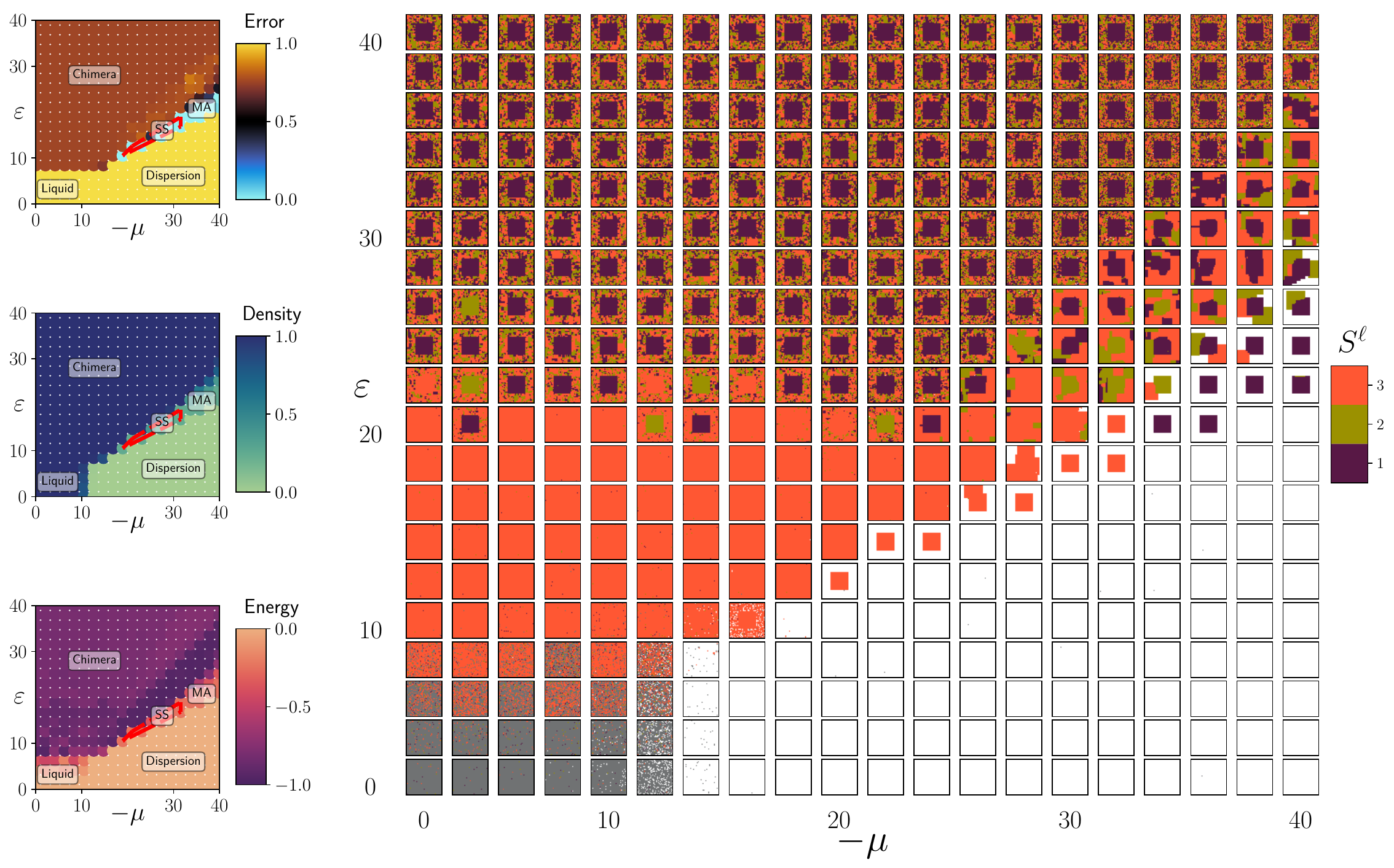}
\caption{Non-equilibrium state diagram using continuous-time simulations, with $\lambda=10$ and shifting sequence $S^1\to S^2\to S^3$. For each value of $(-\mu,\varepsilon)$, a system with $m=3$ stored $40\times 40$ structures is simulated in the $80\times 80$ lattice with $10^8$ reaction steps. Left: Error, density, and energy at the end of the simulations. The different state regions are marked, with MA standing for multifarious assembly, and SS for shape-shifting. The shape-shifting region is also marked by the dashed red lines. Right: Snapshots at the end of the simulations, colored by structure as described in Appendix~\ref{app:Colouring}. 
\label{fig:NESD}}
\end{center}
\end{figure*}

\section{Non-equilibrium multifarious self-organization}

Adding non-reciprocal interactions to the multifarious self-assembly model leads to a system that exhibits shape-shifting between the different target structures, which we visualize by using a simplified coloring scheme described in Appendix \ref{app:Colouring}. 
To identify what state our system is in we use the density of tiles in the lattice, the energy arising from reciprocal bonds being made in the lattice, and the overlaps. Overlaps quantify how much of each structure is currently assembled in the lattice. From the overlaps we can calculate the error, which describes how far the most assembled structure in the lattice is from being fully assembled. In Appendix~\ref{app:orderparams} we provide precise definitions of these three quantities. 

The shape-shifting state is characterized by a low error, with the overlaps for each structure rising and falling along the pre-defined shifting sequence. The dynamics of the overlaps is shown in Fig.~\ref{fig:overlaps}, alongside snapshots of shape-shifting.
In the multifarious self-assembly state, the system can successfully assemble structures (characterized by the low error), but if the non-reciprocal interaction strength $\lambda$ is too low, we do not see shape-shifting.
The chimera state occurs when the reciprocal interaction strength $\varepsilon$ is too high and can be identified by high error, high density, and low energy. The system builds uncontrollably, building unwanted structures attached to the desired ones. The liquid and dispersion states are characterized by a high error and energy, and can be distinguished via the density. The liquid state has a high density, whilst the dispersion state has a low density.

In Fig.~\ref{fig:NESD}, we show the state diagram obtained using continuous-time simulations in non-equilibrium systems, with $\lambda=10$. The different states described are marked in the corresponding locations on the state diagram.
The boundary line for instability towards chimeric structures is the same as that found in equilibrium systems\cite{Metson2025arXiv-a}. In particular, this boundary is different compared with discrete-time simulations, due to the faster chimeric growth dynamics in the continuous-time simulations\cite{Metson2025arXiv-a}.
As found previously in discrete-time simulations\cite{Osat2023NN}, increasing the non-reciprocal interaction strength pushes the liquid-dispersion transition point to larger $-\mu$ and the liquid-chimera transition point to larger $\varepsilon$.

\begin{figure*}[t]
\begin{center}
\includegraphics[width=0.9\linewidth]{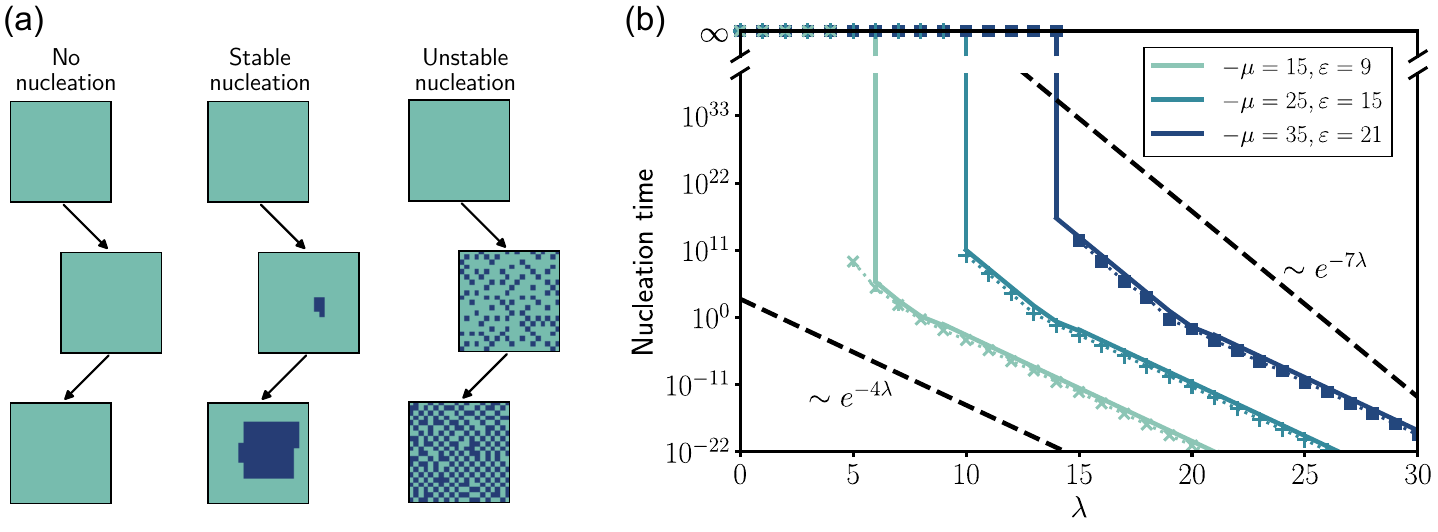}
\caption{Nucleation for shape-shifting. (a) Snapshots of the system in the three different parameter ranges: no nucleation ($\lambda<\frac{2}{3}\varepsilon$), stable nucleation ($\frac{2}{3}\varepsilon<\lambda<\varepsilon$) and unstable nucleation ($\lambda>\varepsilon$). (b) Nucleation time measured in simulations (markers and dotted lines) and analytical results (solid lines). Plotted as a function of $\lambda$ for different values of $(-\mu, \varepsilon)$.
\label{fig:Nucleation}}
\end{center}
\end{figure*}

\section{Mechanisms of shape-shifting}

The inclusion of non-reciprocal interactions in the multifarious self-assembly model brings the system out of equilibrium and leads to shape-shifting behaviors. Depending on the parameter values, we find two different shifting mechanisms. The first is the typical nucleation followed by interface growth, which occurs for $\lambda<\varepsilon$.
The second mechanism occurs when $\lambda>\varepsilon$, and involves the direct replacement of tiles. This is essentially many individual nucleation events which occurs due to the high value of $\lambda$. This is unstable in discrete-time simulations, and can also be unstable in continuous-time simulations, for example with a cyclic shifting sequence.

In the following sections, we decompose shape-shifting into two sub-processes: nucleation and growth. We discuss the mechanisms of each process, and confirm our proposed mechanisms by comparing analytical calculations of the timescales against measurements from simulations of the non-reciprocal multifarious self-organization model.

\subsection{Nucleation}\label{sec:Nucleation}

A shape-shift is initiated by the nucleation of the next structure in the sequence within the structure currently sitting in the system. To investigate the nucleation behavior, we set up the system to store two structures ($S^1$ and $S^2$), with a shifting sequence $S^1\to S^2$, and fill the system with $S^1$. 
Varying $\lambda$ we find three different regimes. Snapshots of the evolution of systems in the three different regimes are shown in Fig.~\ref{fig:Nucleation}(a). For $\lambda<\frac{2}{3}\varepsilon$ we observe no nucleation and $S^1$ remains in the system. We observe canonical stable nucleation for $\frac{2}{3}\varepsilon<\lambda<\varepsilon$. There are many spontaneously created nucleation seeds, and most collapse back to nothing. Eventually however, one of the seeds is able to grow to the critical size, beyond which growing the seed becomes the dominant dynamic pathway of the system and the nucleation is successful. When $\lambda>\varepsilon$, tiles from the next structure arrive individually across the lattice leading to fast but delocalized nucleation. This has a destabilizing effect on the system and does not allow for controlled shifting between structures.

A key quantity of interest is the nucleation time, which we define as the average time required to reach 20\% of $S^2$ in the system.
In Fig.~\ref{fig:Nucleation}(b) we plot the nucleation time as a function of $\lambda$, for different values of $(-\mu,\varepsilon)$. In the plot we show the nucleation time measured in simulations, alongside the analytical results derived in Appendix~\ref{app:nucleation_calc}.

For $\lambda<\frac{2}{3}\varepsilon$, as shown in Appendix~\ref{app:nucleation_calc}, the process of removing a nucleated seed is much faster than the process of growing out the seed, and so in the calculation we consider the nucleation time to be infinite. In this parameter range we do not observe any successful nucleation within the simulation time.

Within the intermediate range $\frac{2}{3}\varepsilon<\lambda<\varepsilon$ we observe stable nucleation. The nucleated region appears within the simulation time and is localized.
In Appendix~\ref{app:nucleation_calc} we calculate the time it takes to nucleate a supercritical seed.
In this parameter range, the time required to grow a supercritical seed is practically the same as the time to reach 20\% of $S^2$ in the system, which is what we measure in simulations. The timescales are similar because the time it takes for a supercritical seed to appear dominates over the time to grow from a supercritical seed to 20\% of the structure. This is because most spontaneous nucleations shrink back to nothing rather than reaching a critical size.
The analytical result
\begin{equation}
    t_{N}\left({\frac{2}{3}\varepsilon<\lambda<\varepsilon}\right)= \frac{1}{L^2}e^{-\mu-7\lambda+5\varepsilon} + e^{-\mu-3\lambda+\varepsilon}, \label{eq:tN:int_maintxt}
\end{equation}
is plotted in Fig.~\ref{fig:Nucleation}(b) and agrees well with nucleation times measured in simulations. 
The first term in the right-hand side of Eq.~\eqref{eq:tN:int_maintxt} is the contribution coming from the time to create new nucleation seeds. For small $\lambda$ this is slow, and so this term dominates. This results in the observed $t_N\sim e^{-7\lambda}$ scaling. As $\lambda$ increases, the time to create nucleation seeds drops, and the second term becomes relevant. As $\lambda$ increases beyond $\varepsilon$, which we discuss in the following paragraph, we see a crossover to $t_N\sim e^{-4\lambda}$ scaling.

\begin{figure*}[t]
\begin{center}
\includegraphics[width=0.9\linewidth]{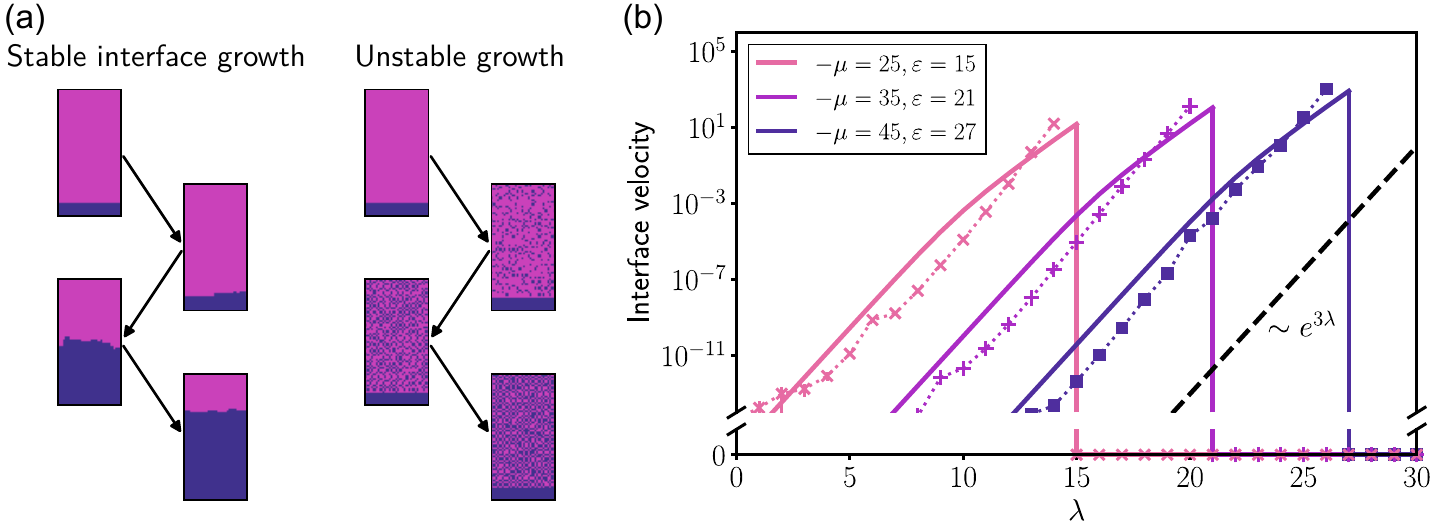}
\caption{(a) Snapshots of stable interface growth ($\lambda<\varepsilon$) and unstable growth proceeding away from the interface ($\lambda>\varepsilon$). (b) Interface velocity measured in simulations (markers) and analytical results (solid lines). For $\lambda>\varepsilon$ growth no longer proceeds via the interface, so we consider the interface velocity to be zero. Plotted as a function of $\lambda$ for different values of $(-\mu, \varepsilon)$.
\label{fig:InterfaceGrowth}}
\end{center}
\end{figure*}

For $\lambda>\varepsilon$, instead of localized nucleation, tiles from the next structure arrive across the system individually. In this parameter region, we can calculate the nucleation time as
\begin{align}
t_{N}({\lambda>\varepsilon}) &\approx f e^{-\mu-4\lambda+2\varepsilon},
\end{align}
where $f<0.5$ is the target fraction of $S^2$ which we set to $f=0.2$. The derivation of this expression is given in Appendix~\ref{app:nucleation_calc}.
In this parameter range shape-shifting is generally not stable, as discussed further in Appendix~\ref{app:nucleation_calc}.

\subsection{Interface growth}\label{sec:interface_growth}

Following nucleation, shape-shifting proceeds via the expansion of the nucleated region. The interface between $S^1$ and $S^2$ grows as the $S^2$ region invades $S^1$ to complete the shape-shift.

To measure interface growth we once again consider a system with two stored structures ($S^1$ and $S^2$) and a shifting sequence $S^1\to S^2$. We initialize the system with a flat interface by placing a slab of $S^2$ underneath a slab of $S^1$. We use hard-wall and periodic boundary conditions perpendicular and parallel to the interface respectively. In order to measure the interface over long times we use interface displacement \cite{Devillard1992E, Osat2024PRL}, which overcomes the issue of a finite system size.
Furthermore, since we are interested in characterizing the behavior of a single interface, when measuring the interface velocity we do not update any tiles further than five vertical lattice sites above the highest point of the interface or five below the lowest point of the interface. This prevents spurious nucleation and the formation of new interfaces \cite{Osat2024PRL}.
On the left-hand side of Fig.~\ref{fig:InterfaceGrowth}(a) we show interface growth in a system with $\lambda<\varepsilon$. In this case the interface moves steadily upwards as the $S^2$ region grows. On the right-hand side of Fig.~\ref{fig:InterfaceGrowth}(a) we show an example of the system evolution when $\lambda>\varepsilon$. As discussed previously, in this case $S^2$ tiles arrive across the whole lattice, independently of the already present $S^2$ region, and we do not see interface growth.

In Fig.~\ref{fig:InterfaceGrowth}(b) we plot the interface velocity as a function of $\lambda$. We show the velocity measured in simulations alongside the analytical results derived in Appendix~\ref{app:interfacegrowth_calc}. The good agreement supports the picture of interface growth used in the calculation.
For $\lambda<\varepsilon$ we observe that the interface velocity scales with the non-reciprocal interaction strength as $v \sim e^{3\lambda}$. This scaling can also be extracted from the analytical result for the interface velocity, which is derived in Appendix~\ref{app:interfacegrowth_calc}
\begin{align}
v_I({\lambda<\varepsilon}) &= \left[\frac{1}{L}e^{-\mu-3\lambda+\varepsilon}
    + \frac{L-2}{2}e^{-\mu-2\lambda}
    + e^{-\mu-\lambda-\varepsilon}\right]^{-1} ,\nonumber \\
&\sim e^{3\lambda}.
\end{align}
This scaling emerges since the rate-limiting step in interface growth is the reaction needed to start growing the next layer, which occurs with a rate that scales as $e^{3\lambda}$.
For $\lambda>\varepsilon$, interface growth is disfavored in comparison to the direct tile replacement mechanism described in the Section~\ref{sec:Nucleation}. This is shown in the snapshots on the right side of Fig.~\ref{fig:InterfaceGrowth}(a). As can be seen in the snapshots, the interface itself does not grow.

\subsection{Discussion}

In discrete-time simulations the timescale of nucleation and interface growth is generally constant with respect to $\lambda$.\cite{Osat2024PRL} This is because the process of proposing a favorable reaction, which involves picking both the correct position and the correct new tile, is a very slow process for moderate to large system sizes.
Therefore, the timescales are not strongly affected by $\lambda$, which only modifies the strength of an already quite high selection probability. 
In contrast, in continuous-time simulations a reaction is always executed, with the most favorable reactions being the most likely reactions to be picked. Furthermore, the time between reactions depends on the total reaction rate, which depends on $\lambda$. Therefore, in continuous-time simulations we see very strong dependencies of the timescales on $\lambda$.
Whilst some systems show similar dynamics between different simulation methods \cite{Branka1998JCP, Strating1999PRE, Whitelam2007TJoCP}, in the system studied here, over the timescales of assembly and shape-shifting we find that the kinetics are not identical between the discrete-time Monte Carlo and the continuous-time Gillespie simulations.
As explained, a key contributor of this is the fact that we are dealing with a multi-component system, with a large number of components. This is true even when detailed balance is obeyed ($\lambda=0$), as explored in the previous work of this series \cite{Metson2025arXiv-a}.
Nonetheless the same key behaviors, including in the non-equilibrium system, are observed using either of the two different simulation methods.


\section{Conclusions}
In this work, we have demonstrated that the non-reciprocal multifarious self-organization model successfully exhibits shape-shifting behavior through a combination of reciprocal and non-reciprocal interactions. By validating this model with continuous-time Gillespie simulations and comparing it to previous discrete-time Monte Carlo simulations, we have uncovered critical insights into the timescales governing shape-shifting. Our simulations revealed two distinct growth mechanisms: interface-driven growth when non-reciprocal interactions are weaker than reciprocal ones, and direct tile replacement when non-reciprocal interactions dominate. These mechanisms provide a deeper understanding of how shape-shifting processes are regulated under different conditions.
The measured nucleation rates and interface velocities agree well with our analytical predictions, reinforcing the accuracy of our theoretical approach. 

Interestingly, we find that the continuous-time Gillespie algorithm leads to different kinetics compared to the discrete-time Monte Carlo algorithm. For example, in the continuous-time simulations the timescales of nucleation and shape-shifting depend strongly on the non-reciprocal interaction strength $\lambda$. In discrete-time simulations the dependence of timescales on $\lambda$ is typically piecewise flat \cite{Osat2023NN, Osat2024PRL}.
Despite these differences, the same fundamental behaviors are observed using either simulation technique.

Extending the model to different types of interactions and environment conditions could broaden its applicability to real-world scenarios, such as adaptive materials or programmable matter.
Aside from creating dynamic states, non-equilibrium strategies have also been exploited to make systems more robust or efficient than their equilibrium counterparts \cite{Behera2023PRX, Navas2024TJoCP, Nag2024JCTC}, including the application of non-reciprocal interactions \cite{Osat2024PRL}. 
A relevant experimental system to our work is that of chemically active phoretic colloids \cite{Soto2014,Soto2015,AgudoCanalejo2019,OuazanReboul2021,OuazanReboul2023b,OuazanReboul2023,OuazanReboul2023a,Grauer2020SR,Ketzetzi2024arXiv,Goth2025CP}. These colloids produce or consume chemicals, and respond to gradients in the resulting chemical fields\cite{Golestanian2022a}.
Mixing colloids with different chemical activities and mobilities generically leads to non-reciprocal interactions. Highly interesting self-organization behaviors have been demonstrated in colloidal systems, both experimentally and in theory. It would be interesting to explore how the dynamics we investigate in this work could be applied to such systems.
Overall, our findings contribute to the growing field of non-equilibrium self-assembly, offering a pathway to design systems capable of dynamic structural reconfiguration.

\begin{acknowledgments}
We acknowledge support from the Max Planck School Matter to Life and the MaxSynBio Consortium which are jointly funded by the Federal Ministry of Education and Research (BMBF) of Germany and the Max Planck Society.
\end{acknowledgments}

\bibliography{refs, refs_JM,Golestanian}

\clearpage

\appendix

\section{Simulations}\label{app:simulations}

\begin{figure*}[t]
\begin{center}
\includegraphics[width=\linewidth]{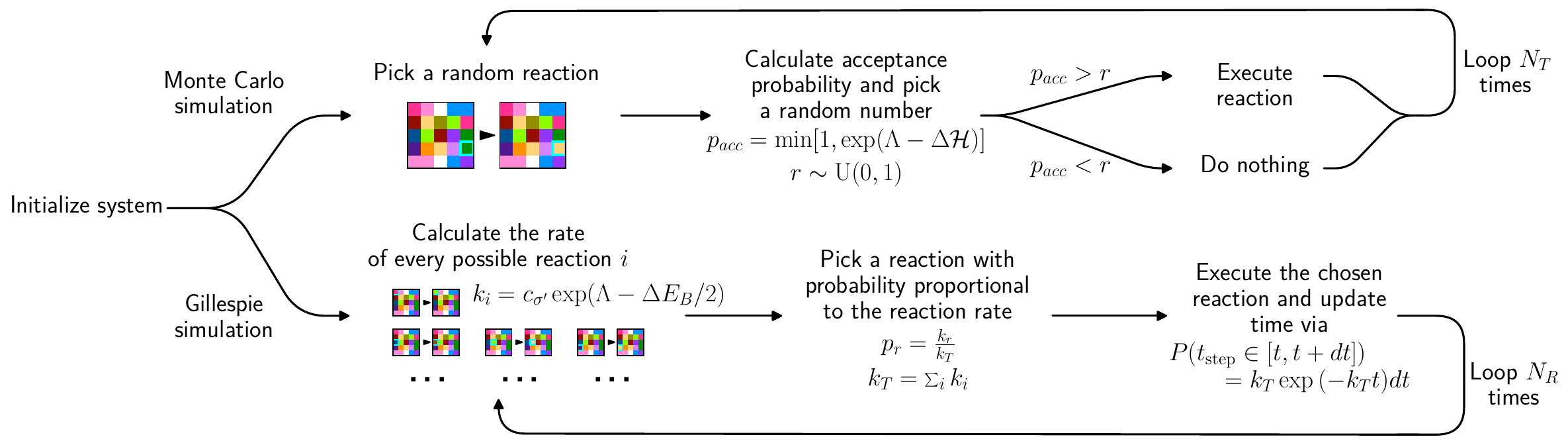}
\caption{Flowchart of the discrete-time Monte Carlo and continuous-time Gillespie simulation algorithms used in this paper to simulate the non-reciprocal multifarious self-organization model. $\mathrm{U}(0,1)$ denotes a uniform distribution over the interval $[0,1]$.
\label{fig:Algorithms}}
\end{center}
\end{figure*}

A schematic of the discrete-time Monte Carlo and continuous-time Gillespie algorithms used in this paper is shown in Fig.~\ref{fig:Algorithms}. The algorithms are the same as those used in the previous work in this series\cite{Metson2025arXiv-a}, except for the addition of the non-reciprocal interactions in the Monte Carlo acceptance probability and in the Gillespie rates.

We use the configuration variable $\sigma_{\alpha}$, denoting the tile at each lattice site $\alpha$, to describe the configuration of the system lattice at any given time. $\sigma_{\alpha}=1, 2, \dots ,M$ represents the different tile species, while $\sigma_{\alpha}=0$ stands for an empty lattice site. We also set $k_{\rm B} T = 1$, and absorb the temperature into the now dimensionless parameters $(-\mu,\varepsilon, \lambda)$.

\subsection{Discrete-time simulations}
At each step in a Monte Carlo simulation\cite{Newman1999}, a random reaction is proposed. In our system, a reaction would involve changing the tile $\sigma_{i,j}$ located at lattice site $\alpha=(i,j)$ to the tile $\sigma'$.
This move can then be either accepted or rejected. The acceptance probability is
\begin{equation}
p_\mathrm{acc} = \min[1, e^{\Lambda-\Delta \mathcal{H}}],
\end{equation}
where $\mathcal{H}$ is defined in Eq.~\eqref{eq:Hamiltonian} and $\Lambda$ is defined in Eq.~\eqref{eq:Lambda}.
If the reaction is accepted then the system is correspondingly updated, otherwise the system remains unchanged. This processes is repeated for the desired simulation time.

\subsection{Continuous-time simulation}
In contrast to accepting or rejecting random moves, in the Gillespie algorithm \cite{Gillespie1976JCP, Gillespie1977JPC, Newman1999} at each step the rate of all possible reactions are calculated. In our case, the rate of changing the tile $\sigma_{i,j}$ located at lattice site $(i,j)$ to the tile $\sigma'$ is
\begin{equation}
    k_{\sigma_{i,j} \rightarrow \sigma'} = c_{\sigma'} e^{\Lambda-\Delta E_B/2}.
\end{equation}
As defined in Section~\ref{sec:Model}, the concentration of species $\sigma'$ in the reservoir is
\begin{equation}
c_{\sigma'}=
\begin{cases}
1, & \sigma' = 0, \\
e^{\mu}, & \text{otherwise}.
\end{cases}
\end{equation}
The change in bond energy is
\begin{multline}
\Delta E_B = 
U^{\rm r}_{\sigma_{i-1,j} \diagup \sigma'}
+ U^{\rm r}_{\sigma' \diagup \sigma_{i+1,j}} 
+ U^{\rm r}_{\sigma' \diagdown \sigma_{i,j-1}} 
+ U^{\rm r}_{\sigma_{i,j+1} \diagdown \sigma'} \\
- U^{\rm r}_{\sigma_{i-1,j} \diagup \sigma_{i,j}} 
- U^{\rm r}_{\sigma_{i,j} \diagup \sigma_{i+1,j}} 
- U^{\rm r}_{\sigma_{i,j} \diagdown \sigma_{i,j-1}} 
- U^{\rm r}_{\sigma_{i,j+1} \diagdown \sigma_{i,j}},
\end{multline}
and $\Lambda$ is defined in Eq.~\eqref{eq:Lambda}.

Once all the rates have been calculated, a random reaction is chosen, with the probability of choosing reaction $r$ being
\begin{equation}
\text{Prob}(\text{selecting } r) = k_r/k_T,
\end{equation}
where $k_T$ is the sum of the rates of all possible reactions. This reaction is executed and the system time is incremented by $t_\text{step}$, which is exponentially distributed with mean $1/k_T$.

\section{Coloring lattice snapshots}\label{app:Colouring}
\begin{figure}[b]
\begin{center}
\includegraphics[width=0.8\linewidth]{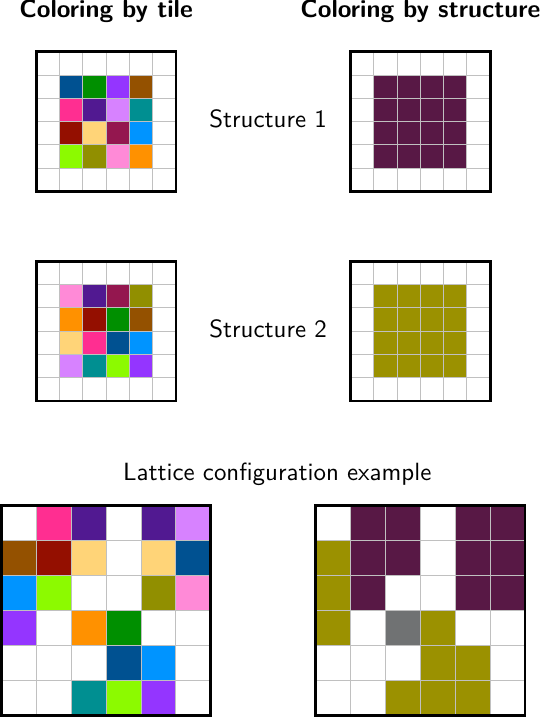}
\caption{An illustration of coloring by tile species and coloring by structure. For clarity we generally color snapshots by structure. White corresponds to empty lattice sites. Tiles which are present in the lattice with no specific interactions are colored gray. All other tiles present are colored based on the specific interactions they make within each structure.
\label{fig:colouring}}
\end{center}
\end{figure}

To display snapshots of the system at a given time in a simulation, one could simply assign a unique color to each tile species, as is done in Fig.~\ref{fig:Cartoon}. However, for large structures defined as random arrangements of tiles, this quickly becomes hard to visually interpret. Therefore we color snapshots based on the different structures, or parts of structures, assembled in the lattice. First we assign a unique color to each of the $m$ structures stored in the system.
Empty lattice sites are colored white. Tiles which are present in the lattice with no specific interactions are colored gray. For each remaining tile, we count through its neighbors in the lattice and assign it the color of the structure for which it has the most neighbors currently in the lattice.
An illustration of the coloring by tile and coloring by structure schemes is shown in Fig.~\ref{fig:colouring}.

When the structures are the same size as the lattice, we employ a simpler coloring algorithm. Each tile in the lattice inherits the color of the structure with a tile at that lattice point. In the rare event that a tile in the lattice shares the same position to more than one structure, we pick randomly from the colors of the relevant structures.

\section{Order parameters} \label{app:orderparams}
To characterize what state our system is in, we use the overlaps, density and bond energy.

To find the overlaps, we first find the largest connected cluster of tiles $G$. For each structure we then calculate $A_\ell=G \cup S^{\ell}$, where $S^{\ell}$ is $\ell$th structure located at the center of the lattice. The overlap for each structure is then given by 
\begin{equation}
O_\ell={|A \cap S^\mathrm{\ell}|}/{|A|}. \label{eq:overlaps}
\end{equation}
The overlaps can be used to distinguish the shape-shifting state. In the shape-shifting state the overlaps rise and fall according to the predefined shifting sequence as can be seen in Fig.~\ref{fig:overlaps}(b).
From the overlaps we can also calculate the error, which is given by
\begin{equation}
\text{Error} = 1-O,
\end{equation}
where $O = \max_\ell [O_\ell]$.
The error ranges between 0 and 1.

The density is the fraction of the lattice filled with tiles,
\begin{equation}
\rho = \frac{1}{L^2}\sum_{\alpha} (1-\delta_{0,\sigma_\alpha}),
\end{equation}
which also ranges between 0 and 1.

We normalize the energy with respect to the maximum bond energy $E_\mathrm{max}$, such that the energy takes values between 0 and -1, 
\begin{equation}
E = \frac{\sum_{\langle \alpha, \beta \rangle} U^{\rm r}_{\sigma_\alpha \square \sigma_\beta}}{E_\mathrm{max}}.
\end{equation}

\section{Analytical timescale calculations}
Here we present the analytical calculations we make of the different timescales of shape-shifting. The calculation method reflects the dynamics of the Gillespie algorithm and involves calculating the rates of all the dominant transitions given a particular lattice configuration. These can then be used to calculate the probability of transitioning to the different subsequent states as well as the average time required to do so.

Our calculation technique relies on several simplifying assumptions which make the calculation tractable by hand. Firstly we 
focus only on structural identity, without needing to identify each particular tile species. This essentially neglects the chance that a tile shares neighbors in multiple structures, which becomes vanishingly small in the limit of large $l$. 
We also neglect transitions involving empty sites, assuming that these transitions are subdominant. This can be seen, for example, in the snapshots shown in Fig.~\ref{fig:Nucleation}(a) and Fig.~\ref{fig:InterfaceGrowth}(a), in which there are no empty lattice sites.
Typically a few key transitions will have by far the largest rate, such that the average time for that transition is approximately the inverse of the sum of the rates of these dominant transitions. 
For different parameter values, different transitions dominate. Based on this, we divide the calculations of timescales into different parameter ranges.
These assumptions greatly simplify the calculations, as a comprehensive enumeration might not be tractable. Despite these assumptions, the resulting calculations provide a good approximation. The main reason for this is the exponentiation in the rates. This means that particular transitions quickly become very dominant, which can be used to simplify rate calculations. Furthermore, for each process there is typically one rate-limiting step (RLS) which dominates the timescales \cite{Lifshitz1961JoPaCoS}. In nucleation the RLS is forming a critical seed (reactions $\mathcal{A}_N\to\mathcal{B}_N\to\mathcal{B}_N\to\mathcal{C}_N$ in Fig.~\ref{fig:Calc_Nucleation}). In interface growth the RLS is initiating a new layer (reaction $\mathcal{A}_I$ in Fig.~\ref{fig:Calc_Interface}). This means that the scaling in the timescales is typically due to one key process, and once this process has been considered in the calculation, the correct scaling will emerge.

\subsection{Nucleation calculation}\label{app:nucleation_calc}

\begin{figure}[t]
\begin{center}
\includegraphics[width=0.7\linewidth]{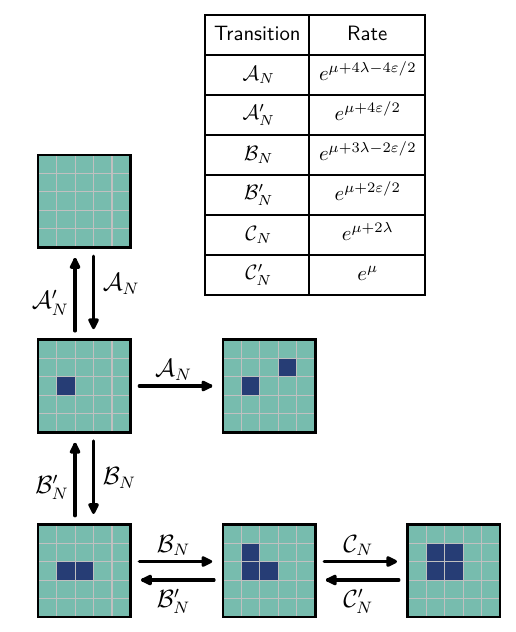}
\caption{Important individual reaction transitions used in the calculation of nucleation time.
\label{fig:Calc_Nucleation}}
\end{center}
\end{figure}

In this section we calculate the time to nucleate structure $S^2$ inside $S^1$, in a system with periodic boundary conditions and a shifting sequence $S^1\to S^2$. Depending on the value of $\lambda$, the dynamics are qualitatively very different, so we split the calculation up into three parts, each valid for a particular range of $\lambda$. The important transitions for the calculations are shown in Fig.~\ref{fig:Calc_Nucleation}.

Below $\lambda<\frac{2}{3}\varepsilon$ nucleation is not observed. To understand this, consider the growth of a small seed. Continuing the growth of the seed depends on the transition $\mathcal{B}_N$ being more favorable than the reverse transition $\mathcal{B}'_N$. The ratio of the rates of these transitions is
\begin{equation}
    \frac{\R_{\mathcal{B}_N}}{\R_{\mathcal{B}'_N}} = e^{3\lambda - 2\varepsilon}.
\end{equation}
For $\lambda<\frac{2}{3}\varepsilon$ this ratio is less than 1, meaning the growth of seeds is heavily disfavored, such that we can consider the rate of nucleation to be practically zero.

For $\lambda>\varepsilon$ the transition that dominates is $\mathcal{A}_N$, which has rate $\R_{\mathcal{A}_N}=e^{\mu+4\lambda-4\varepsilon/2}$. The dynamics evolve via many individual nucleations of isolated $S^2$ tiles. The time to reach some small fraction $f$ of $S^2$ in the system is thus given by 
\begin{align}
t_{N}({\lambda>\varepsilon}) &\approx fL^2 \frac{1}{L^2 \R_{\mathcal{A}_N}+\cdots}, \\ 
  &\approx \frac{f}{\R_{\mathcal{A}_N}},\\
  &\sim e^{-4\lambda}. \label{eq:tN:lge}
\end{align}
The dots in the first line represent the contribution to the total rate at each step from other transitions. In this parameter range the other transitions have a negligible rate compared to the $\mathcal{A}_N$ transition.
In discrete-time Monte Carlo simulations this parameter range is always unstable \cite{Osat2023NN}, destroying any structure in the system. In continuous-time Gillespie simulations, a shifting sequence $S^1\to S^2$, can successfully shape-shift.
This is because the rates of the $\mathcal{A}_N$ transitions, and subsequent filling-in of the checkerboard pattern, are fast enough to shape-shift before the structure becomes unstable.
With any other shifting sequence the Gillespie simulations do not lead to successful shape-shifting, because subsequent shifts start too quickly.

Between $\frac{2}{3}\varepsilon<\lambda<\varepsilon$ we observe stable nucleation, consisting of the creation of small seeds, many of which are destroyed. Eventually however, one seed is able to grow to a critical size, beyond which growth is fast. In the initial state the transition $\mathcal{A}_N$ is the only relevant transition. Following the $\mathcal{A}_N$ transition, for $\lambda<\varepsilon$ the system most often replaces the newly added $S^2$ tile with the old $S^1$ tile, going back to the very stable initial state via the $\mathcal{A}'_N$ transition. However, after many nucleation attempts, eventually transition $\mathcal{B}_N$ is picked over $\mathcal{A}'_N$. On average this takes around $\mathcal{A}'_N/\mathcal{B}_N$ attempts. After the $\mathcal{B}_N$ transition, the remaining transitions to grow out of the seed are dominant, since $\lambda>\frac{2}{3}\varepsilon$. Based on this, we can calculate the nucleation time to be
\begin{align}
t_{N}\left({\frac{2}{3}\varepsilon<\lambda<\varepsilon}\right) &= \frac{\R_{\mathcal{A}'_N}}{\R_{\mathcal{B}_N}} \left( \frac{1}{L^2 \R_{\mathcal{A}_N}} + \frac{1}{\R_{\mathcal{A}'_N}} \right) + \cdots \\
  &= \frac{1}{L^2}e^{-\mu-7\lambda+5\varepsilon} + e^{-\mu-3\lambda+\varepsilon}.
\end{align}
Here the dots in the first line represent the time to grow after a seed has successfully reached the critical point, which is negligible compared to the time required to get a seed to the critical point. 


\subsection{Interface growth calculation}\label{app:interfacegrowth_calc}
\begin{figure}[t]
\begin{center}
\includegraphics[width=0.7\linewidth]{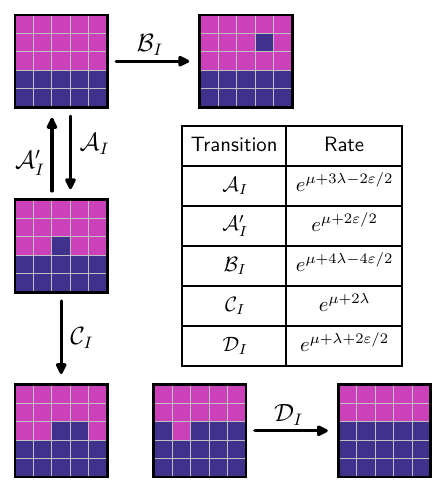}
\caption{Important reaction transitions used in the calculation of interface growth velocity.  
\label{fig:Calc_Interface}}
\end{center}
\end{figure}

In this section we calculate the velocity of an interface between two structures $S^1$ and $S^2$, in a system with a shifting sequence $S^1\to S^2$ and periodic boundary conditions in the direction parallel to the interface. The important transitions for the calculation of interface velocity are shown in Fig.~\ref{fig:Calc_Interface}.

To calculate the interface velocity, we start by calculating the average time taken to grow a full layer. Consider an initially flat interface. The first step is to nucleate the next layer, which is done in the transition labeled $\mathcal{A}_I$. The rate of the reaction to do this is $\R_{\mathcal{A}_I}=e^{\mu+3\lambda-2\varepsilon/2}$. For $\lambda<\varepsilon$ this transition dominates over the $\mathcal{B}_I$ transition, and so the average time for this step is $t_{\mathcal{A}_I}\approx(L\R_{\mathcal{A}_I})^{-1}$. The factor of $L$ is present since there are $L$ different sites along the interface where this reaction could occur. Now the dominant transitions are to grow out this new layer, using transitions like $\mathcal{C}_I$. This involves $L-2$ reactions, each of which takes on average a time $t_{\mathcal{C}_I}\approx(2e^{\mu+2\lambda})^{-1}$. Adding the final $S^2$ tile to the new layer has a different rate, since the site is now surrounded by two $S^2$ tiles. This step takes a time $t_{\mathcal{D}_I}\approx(e^{\mu+2\lambda+2\varepsilon/2})^{-1}$. Putting this all together, the total average time to grow a new layer is
\begin{equation}
    t_\mathrm{layer} \approx \frac{1}{L}e^{-\mu-3\lambda+2\varepsilon/2}
                   + \frac{L-2}{2}e^{-\mu-2\lambda}
                   + e^{-\mu-\lambda-2\varepsilon/2}.
\end{equation}
To get the interface velocity, we simply invert the time taken to grow a single layer which gives
\begin{equation}
    v \approx \left[\frac{1}{L}e^{-\mu-3\lambda+2\varepsilon/2}
                   + \frac{L-2}{2}e^{-\mu-2\lambda}
                   + e^{-\mu-\lambda-2\varepsilon/2}\right]^{-1}.
\end{equation}
In practice a second layer can nucleate before the first one has finished growing - particularly for large interfaces and for $\lambda$ values closer to $\varepsilon$. These additional reactions are not included in the calculation, in the spirit of only considering the most dominant reactions.

To identify the upper bound in $\lambda$, one can see that for $\lambda>\varepsilon$, direct replacement of $S^2$ tiles in the bulk of $S^1$ (transition $\mathcal{B}_I$) becomes the dominant reaction, as discussed in the calculation of the nucleation time. This leads to the breakdown of interface growth, and so we consider the interface velocity to be zero in this parameter range.

\end{document}